A Chorus of Bells

Jeremy Bernstein

Stevens Institute of Technology

"I did not dare to think that it was false, but I **knew** it was **rotten!**" John Bell

Not long after he matriculated at Queens College in Belfast in 1945, John Bell took his first course in quantum mechanics from Robert Sloane. At the time Bell had vivid red hair but not the beard he wore later when he scarred his lip in a motorbike accident. One pities poor Sloane. Most students when they first encounter quantum mechanics are in a state of shock and awe. Not Bell. He decided that at its base it was fraudulent. He had screaming arguments with Sloane. Of course then, and thereafter, Bell accepted all the practical applications of quantum mechanics. He later introduced the acronym FAPP-For All Practical Purposes. He agreed that quantum mechanics was the greatest FAPP theory ever created. He was always sure that it would pass the various tests he proposed for it. But it was the muddle that he perceived in its foundations he could not stand. Take the wave function for example.

When we learn, to take an example, about the quantum mechanics of the electron in the hydrogen atom, we have, I am sure, some sort of picture of a tiny charged object whose position is described by its wave function. All of our instincts tell us that the electron **has** a position which the wave function is telling us about. We must keep reminding ourselves that if we believe the interpretation of the quantum theory as expressed say by Bohr then the wave function is not a description of reality. It **is** reality. As Bohr put it,
"There is no quantum world. There is only an abstract quantum physical description. It is wrong to think that the task of physics is to find out how nature is .Physics concerns only what we can say about nature."[1] Bell found this totally

---
[1] This is quoted and discussed in The Philosophy of Quantum Mechanics by Max Jammer, John Wiley, New York, 1974,p 204.It is actually something that Bohr's assistant at the time Aage

unacceptable. Even more unacceptable did he find what quantum theory-at least the usual interpretation-had to say about measurement.

In the theory there are "observables" represented by self-adjoint operators. These operators have real eigen-values and associated eigen-vectors. If the system is in a state ψ, and the observable in question is A, then we can expand ψ in a sum over the eigen-vectors associated with A. The coefficients in the sum are complex numbers whose absolute squares represent the relative probabilities of measuring given eigen-values.[2] This is an assumption which is often called "Born's rule" after Max Born who introduced the probability interpretation of the quantum theory. Bohr insisted that there were "apparatus" and that these were necessarily described by classical; ie, non-quantum, physics. These apparatus performed measurements on quantum systems. He was never very clear exactly how to make this distinction except that systems were "small" and the apparatus were "large." This lack of precision drove Bell crazy and he kept referring to Bohr as an "obscurantist." FAPP there was in general no problem and it is a separation that experimental physicists make on a daily basis. We could, of course, insist that an apparatus was as quantum mechanical as anything else. But then we are apparently driven into an infinite regress ending up with the experimenter's brain. On top of this there was the act of measurement itself. An actual measurement projects out one of the components of the wave function, something that cannot be described using the formalism of the quantum theory that applies to the behavior of the system up to the time when this measurement is actually recorded. What are the dynamics of this collapse? When exactly does it take place and does it require the consciousness of an "observer" to make it happen? It was over matters like this, where he had his screaming arguments with poor Doctor Sloane.

Bell was philosophically inclined even in high school. He used to bring home from the library large books of Greek philosophy His working class

---

Peterson reported Bohr as having said. For a delightful account of what Bohr did and did not say, see "What's Wrong With This Quantum World" by N.David Mermin, Physics Today, February 2004, 10-11. Bohr said a great many things only some of which are comprehensible to me.
[2] The state vector is normalized to unity which permits this interpretation.

parents referred to him as "the professor"-little did they know. In 1948, Born delivered the so-called Waynflete Lectures at Oxford. Soon after they were published under the title *Natural Philosophy of Cause and Chance*[3]. Bell was much taken by the lectures. However he came across the following,

"I expect that our present theory will be profoundly modified. For it is full of difficulties which I have not mentioned at all-the self-energies of particles in interaction, and many other quantities, like collision cross-sections lead to divergent integrals. But I should never expect that these difficulties could be solved by a return to classical concepts. I expect just the opposite, that we shall have to sacrifice some current ideas and use still more abstract methods. A more concrete contribution to this question has been made by J.v.Neumann in his brilliant book *Mathematische Grundlagen der Quantenmechanik*. He puts the theory on an axiomatic basis by deriving it from a few postulates of a very plausible and general character about the properties of 'expectation values' (averages) and their representations by mathematical symbols. The result is that the formulation of quantum mechanics is uniquely determined by these axioms; in particular no concealed parameters [hidden variables] can be introduced with the help of which the indeterministic description could be transformed into a deterministic one…"[4]

But in early 1952 the papers of David Bohm appeared. Bohm had revived an approach to the quantum theory that had first been introduced by Louis de Broglie in the late 1920s. De Broglie considered the Schrödinger wave function as describing a "pilot wave" that guided the motion of some more or less classical particles. At a meeting at which de Broglie described his scheme he was subjected to withering criticism by Pauli and de Broglie dropped the subject. It was discovered independently by Bohm some three decades later. Bohm found no difficulty in dispatching Pauli's objections. Indeed Bohm's formalism, which I will discuss shortly, can reproduce all the results of non-relativistic quantum theory in a deterministic fashion and hence is a prima facie counter example to von Neumann's claim. When Bell saw this he realized that something had to have been wrong with von Neumann. By this time Bell had graduated with first-class honors from Queen's and had gone to work at a sub-station of the Atomic Energy Research Establishment at Malvern in Worcestershire. He was assigned to work on the design of a linear accelerator. Up to this point there had been nothing he could do about von Neumann since Bell did not read German and von Neumann's book had not yet been translated into English. But at Malvern he found a colleague named Fritz Mandl who both knew German and was interested in the foundations of the quantum theory. He translated the relevant parts of von Neumann.

---

[3] A more recent edition is Dover Publications, New York, 1964.
[4] Born, 1964, op.cit. p.109.

I have read this section of von Neumann several times and each time I am amazed that Bell could extract with such clarity the central point. Incidentally, Basil Hiley who was a close collaborator of Bohm's informs me that he and Bohm "puzzled of over von Neumann for a considerable time but could not spot where the problem lay."[5] Von Neumann was mathematician and a very great one. His book with its axioms and theorems reads more like a math text than a book about physics. There is to be sure physics. For example he presents the first accurate description of the measurement process in the quantum theory. The discussion of what von Neumann refers to as "hidden variables" appears unexpectedly towards the end of the book.[6] To understand it I will remind the reader that if the state of a system is described by a wave function $\varphi$ then the "expectation value" of an observable A, <A> is given by

$$<A> = \int \varphi^* A \varphi dV.$$

In terms of this expectation value the square of the "dispersion" of this observable in this state is given by

$$(\Delta A)^2 = <A^2> - <A>^2.$$

Von Neumann's way of formulating the hidden variable problem is to consider what he calls "dispersion free" states, for which the above quantity is zero. If $\varphi$ happened to be one of the eigen vectors of A, then as far as A was concerned

---

[5] I thank Basil Hiley for this and for other comments.
[6] More exactly on page 320 of the English Edition, Mathematical Foundations of Quantum Mechanics, Princeton University Press, Princeton, 1955.

this state would be dispersion free. Von Neumann proposed taking an ensemble of such states and averaging over them somehow to reproduce the results of quantum mechanics. He then argued that this is impossible. "There are no ensembles free of dispersion," he writes.[7] The assumption he makes-for standard quantum mechanics it is a trivial consequence of the definition of the expectation value- is that expectation values are linear;ie,

    <αA+βB>=α<A>+β<B>

even if A and B do not commute which is a remarkable result if one thinks about it. But eigen-values of sums of non-commuting operators are not additive. Bell's favorite example involves the Pauli spin matrices. The eigen values of

$\sigma_x = \begin{pmatrix} 0 & 1 \\ 1 & 0 \end{pmatrix}$ are ±1 as are the eigen values of $\sigma_y = \begin{pmatrix} 0 & -i \\ i & 0 \end{pmatrix}$ while the eigen-values of the sum are ±√2. But in a dispersion free state the expectation value of an observable must equal one of its eigen values which is not true here since the eigen values are not additive and the expectation values are. This is certainly correct and knocks down the straw hidden variable theories that von Neumann considers, but it has absolutely nothing to do with the de Broglie-Bohm mechanics. I will henceforth refer to this as Bohmian mechanics since I will be using his formalism. I am aware of the fact that he did not like this terminology but it is in common use.

    In this mechanics there are particles that follow classical trajectories which are determined by first order differential equations for the particle coordinates **X**(t). I will begin by considering a single particle. As we shall see, what drives the differential equation-the "force" term- is a wave function ψ(**x**,t) where **x** is any point In space including **X**. ψ satisfies the Schrödinger equation

    i∂/∂tψ(**x**,t)=Hψ(**x**,t).

Here H is the Hamiltonian may include a potential V(**x**). To write the equation for **X**(t) we introduce the current **J**(x,t)

    **J(x,t)**=1/2im(ψ*(**x**,t)∂ψ(**x**,t)-ψ(**x**,t)∂ψ*(**x**,t))

---

[7] Von Neumann, op.cit. p. 323.

where m is the mass of the particle. We also introduce the density $\rho(\mathbf{x},t)$ where

$$\rho(\mathbf{x},t)=\psi^*(\mathbf{x},t)\psi(\mathbf{x},t).$$

Using the Schrödinger equation one can establish the continuity equation

$$\partial/\partial t \rho + \partial \cdot \mathbf{J}=0.$$

The equation for the trajectory of $\mathbf{X}(t)$ is given by-an assumption

$$d\mathbf{X}(t)/dt=\mathbf{J}(\mathbf{X}(t),t)/\rho(\mathbf{X}(t),t)$$

It is comforting to report that for a free particle, V=O,

$$d\mathbf{X}(t)/dt=\mathbf{p}/m.$$

Incidentally, Bohmian mechanics is very often called a "hidden variable" theory. It seems to me that this is a misnomer. There is nothing hidden about the position variables of the particles. It would I think be better to call it a "classical variable" theory. The quantum mechanical features enter because while, given a set of initial conditions the trajectory is then determined, these initial conditions are distributed with probabilities given by $|\psi(\mathbf{x},o)|^2$. Many examples have been worked out including the notorious double slit experiment. In Bohmian mechanics the particle goes through one slit or the other while the guide wave goes through both which accounts for the interference pattern.

While Bohm does discuss the "non-locality" of the theory it was Bell who first stated this feature with clarity. I find a good deal of confusion in discussions of this so I am going to introduce the notions of "strong" and "weak" non-locality. I begin with strong non-locality. I will define this by saying that a theory that is strongly non-local has "tachyons"-particles that always move faster than light –in it. I am well aware that people who discuss this kind of non-locality often mention super-luminal signals that transport "information." This brings in a discussion of what a signal is and what information is that I want to avoid. It is well-known that tachyon theories can be made Lorentz invariant. That is not the problem. The problem is with causality. This difficulty has been known since Einstein first pointed it out in 1907.[8] If there is a faster than light particle that propagates between two space time points with the absorption event occurring later in some reference system, than the emission, then it is possible to find a Lorentz transformation to a system moving less than the speed of light in which the order of these events is reversed. We would now say that in this system the absorption of the tachyon has been converted into the emission of an anti-tachyon. We can play all sorts of games with this. Bell even invented the perfect tachyon murder.[9] The perpetrator shoots the victim in one coordinate system, but to the jury in another system it looks as if an anti-tachyon has been emitted followed by the demise of the victim-no murder.

Tachyons are undesirable and Bohmian mechanics does not have them. But there is weak non-locality which is an ineluctable feature of the quantum theory. Einstein referred to it as "spooky actions at a distance" and Schrödinger coined the term "entanglement" In the paper in which he introduced the term he wrote,

:

When two systems, of which we know the states by their respective representatives, enter into temporary physical interaction due to known forces between them, and when after a time of mutual influence the systems separate again, then they can no longer be described in the same way as before, viz. by endowing each of them with a representative of its own. I would not call that *one* but rather *the* characteristic trait of quantum mechanics, the one that enforces its entire departure from classical lines of thought. By the interaction the two representatives [the quantum states] have become entangled. [10]

It is clear that any scheme that purports to reproduce the quantum theory must have this feature which I have called weak non-locality. Bohmian mechanics does have it. This shows up when we have two particles in an interaction which has produced an entangled state. Each particle has its own

---

[8] *Ann Phys.Lpz.*,**23** ,(1907) 371.
[9] J.S.Bell, Speakable and Unspeakable in Quantum Mechanics,Cambridge University Press, New York, 2004 p.235-6.
[10] E. Schrödinger, Discussion of Probability Relations Between Separated Systems, Proceedings of the Cambridge Philosophical Society, **31**, 1935, 555.

differential equation driven by a common wave function. But if the particles are entangled this wave function $\psi(\mathbf{x}_1,\mathbf{x}_2,t)$ is not separable. The time t is common because the theory is non-relativistic. Hence the behavior of one of the particles is dependant on the instantaneous behavior of the other however widely separated. There are no tachyons here, but just entanglement. In 1966 Bell published in *The Reviews of Modern Physics* an article entitled "On the problem of hidden variables in quantum mechanics."[11] He ends it by saying, "It would be interesting perhaps to pursue some further "impossibility proofs" replacing the arbitrary axioms objected to above by some conditions of locality or of separability of distant systems." But to this there is attached a footnote which he added in proof, that this work had at the time of the publication of this article already been done. This is of course a reference to the inequality that he had derived,

                Rigorous proofs of this inequality abound and I have no intention of reproducing any of them. Instead I am going to give a poor man's version which in its outlines was suggested to me by Bell when I asked him how he explained it to non-specialists with a limited attention span. I have "gussied up" Bell's version. I do this by introducing what I call "Einstein robots." These are incredibly smart robots that can be programmed to reproduce the results of quantum mechanics. They can be made so small that they can fit on single atoms. The one thing they cannot do is to exchange signals of any kind faster that the speed of light. No tachyon guns for them. I am going to program the robots to reproduce the Stern-Gerlach experiment. You will recall that in 1922 Otto Stern and Walther Gerlach sent beams of silver atoms through an inhomgeneous magnetic field. Much to their surprise the beam was split in two and produced two separated lines on a photographic plate. What they did not know at the time was that they had measured the spin of the electron. On the one hand the notion of spin had not yet been invented. On the other hand the electronic structure of silver was not yet known. The core of the silver electrons

---

[11] This article is reprinted in Bell 2004, op cit. The page numbers I will cite are from this reference in this instance page 11.

are in a net state of zero angular momentum while a single valence electron in an S state is outside. This electron spin gives the atom its net angular momentum.

The silver atoms with their attached robots are launched in a beam with a random mixture of spin "up" and spin "down" atoms. When an atom comes under the influence of the inhomogenous magnetic field there is a force on it whose direction depends on the orientation of the spin. When the robot senses this direction it guides the atom along the appropriate orbit. This way the Stern-Gerlach pattern is reproduced. Having accomplished this with no difficulty the robot is given a new task. Now there are two magnets one behind the other. The robot collects all the spin up events from the first magnet and guides them to the second magnet. If its field is oriented in the same direction as the first, the robot will guide all the atoms in the spin up direction. But suppose we rotate the second magnet around the direction of the incoming beam and through an angle θ. Quantum mechanics tells us[12] that with this rotation if the spin was up in the original system then the probabilty of finding it up in the rotated system is $\cos(\theta/2)^2$ while the probability of finding is down is $\sin(\theta/2)^2$. Hence with this rotation there will now be two lines on the photographic plate with varying intensities. When the two magnets are at right angles the intensities are the same. All of this we can teach to the robots.

Now we give the robots a new and more interesting task. We prepare two silver atoms in a spin singlet state whose wave function is symbolically

$(\uparrow_1\downarrow_2 - \downarrow_1\uparrow_2)/\sqrt{2}$ where the arrows refer to the directions of the spin. This is the canonical example of an entangled state. The silver atoms fly off in opposite directions with their robots attached. They encounter two widely serparated Stern-Gerlach magnets. Each robot is on its own and guides its silver atom depending on the orientation of the magnets as it has been instructed to do. If the magnets are parallel the anti-correlation of the two spins is observed. If one of the magnets is rotated through an angle ±θ then one of the robot can be instructed for a fraction of the time proportional to $\sin(\theta/2)^2$ to guide the trajectory of the silver atom so that the two spins are measured to be in the same direction. This agrees with the quantum mechanical result. (See the appendix for the details.) But suppose one magnet is rotated through θ and the other through –θ. Each robot will act as if it is supposed to change its orbit a fraction of the time proportional to $\sin(\theta/2)^2$ so according to the robots the total fraction of the time when the two spins are measured to be the same is proportional to $2\sin(\theta/2)^2$. But the correct quantum mechanical result is $\sin^2(\theta)$ so we are stuck. In the interval 0<θ<π/2 we have $\sin^2(\theta) > 2\sin(\theta/2)^2$. This is a primitive example of a Bell inequality. [13] Quite generally no local hidden variable theory can reproduce all the results of quantum mechanics.

---

[12] See the appendix for the details.

Having spoken to Bell about all of this, I am quite sure that he believed that any experiments done on his inequalities would agree with quantum mechanics. Quantum mechanics gives correct results in domains as widely separated as super-conductivity and super-novae. It would be somewhat absurd to think that it would break down in a Stern-Gerlach experiment and indeed it

---

[13] The purpose of this footnote is to remind the reader of, or introduce the reader to Bell's original inequality which he published in Physics **1** (1964) 195-200. The context is again a double Stern-Gerlach experiment , Let **a** be the direction of one magnet and **b** the direction of the other. Let λ be some "hidden variable." There might be several but one will do. The result of a measurement with the A magnet is given by A(**a**,λ)=±1 while the result of a measurement with the B magnet is B(**b**,λ)=±1. The locality is represented by the fact that A is only a function of **a** and B is only a function of **b**. The correlation of these measurements is given by a function P(**a**,**b**) which a weighted integral over λ with a weight function ρ(λ);ie,

$$P(\mathbf{a},\mathbf{b}) = \int \rho(\lambda) A(\mathbf{a},\lambda) B(\mathbf{b},\lambda) d\lambda.$$

The quantum mechanical result which is derived in the appendix is given by

$$P(\mathbf{a},\mathbf{b})_{qm} = -\cos(\mathbf{a}\cdot\mathbf{b}).$$

Bell asked is it possible to reproduce this answer with any choice of the functions that enter the integral above. Bell derived the inequality below where **c** is a third direction

$$1 + P(\mathbf{b}\cdot\mathbf{c}) \geq |P(\mathbf{a},\mathbf{b}) - P(\mathbf{a},\mathbf{c})|$$

He showed that P(**a**,**b**)$_{qm}$ cannot satisfy this inequality for all choices of direction.

didn't when it was tested by people like Alain Aspect. Bell once said to me with some regret that it showed that Einstein was wrong and Bohr was right. Einstein, he felt, was acting like a reasonable scientist while Bohr was an obscurantist. "The reasonable thing," he said, " just doesn't work." I do not fully understand what Einstein wanted. As a guess I think he wanted to see quantum mechanics emerge from some underlying deterministic theory in somewhat the same sense that thermodynamics emerges from statistical mechanics. He no doubt wanted the underlying theory to be local, free of spooky actions at a distance. What Bell showed is that the underlying theory, if there is one, cannot be local. We know Einstein's feelings about Bohmian mechanics. He expressed them in a letter to Born dated May 12, 1952

"Have you noticed that Bohm believes (as de Broglie did, by the way 25 years ago) that he is able to interpret the quantum theory in deterministic terms? That way seems too cheap to me. But you, of course, can judge this better than I."[14]

I wish I knew what Einstein meant by "cheap" in this context.

When I began to learn quantum mechanics around 1950 there were not that many texts available. One of the standard ones was *Quantum Mechanics* by Leonard Schiff. It was essentially a more detailed write up of the lectures Robert Oppenheimer had given for many years at Berkeley and CalTech. It is a good text from which to learn how to solve problems, but there is nothing concerning what we would now call the foundations of the theory. The same thing is true of Dirac's masterful *The Principles of Quantum Mechanics.* In the first chapter he states that a measurement collapses the wave function and that is that. He once remarked to someone that he thought that it was a good book but that the first chapter was missing. But in 1951 Bohm published his text *Quantum Theory*[15] It is full of discussion of the foundations. Abner Shimony, who made very basic contributions to the development of Bell's inequalities, asked his then thesis advisor Eugene Wigner what he thought of the book. Wigner told him that it was a good book except that there was too much "schmoozing." The schmoozing is just what I liked since it dealt with the foundations of the theory. What is remarkable about the book is that it contains a "proof" that the results of the quantum theory cannot emerge from hidden variables. He writes "We conclude that no theory of mechanically determined hidden variables can lead to *all* of the results of the quantum theory." But not long after the book was published he had produced a theory which did just that. One of the things that Bell took from the book was Bohm's novel presentation of the Einstein, Podolsky, Rosen experiment which they first published in 1935,[16] This version of the EPR experiment has been with us every since. The ingredients will be familiar.

---

[14] The Born-Einstein Letters, Walker and Company, New York, 1971, p. 192.
[15] Prentice Hall, Englewood, New Jersey,

Some mechanism produces a pair of spin-1/2 particles in a singlet state. They fly off in opposite directions to a pair of Stern-Gerlach magnets. Let us say that one of the magnets is oriented in the z-direction and let us say that it measures the spin of one of the particles to be "up." Because of the correlation we have already discussed we would predict that, when measured, the spin of the other particle will be "down." EPR go a step further. They would argue that in this set up the z-component of the spin of the other particle has been implicitly measured and that this implicit measurement has conferred "reality" on this quantity. One can then set about to measure the x-component by rotating the magnet. This having been done we have both components measured which quantum mechanics says is impossible. The solution to this problem, if it is a problem, is to insist that "implicit measurements" in the quantum theory don't count. Either you measure something or you don't. You cannot measure the x and z components simultaneously. You need two different experiments. Bell of course understood this, but I think that it was thinking about double Stern-Gerlach experiments in this context that set him off.

In the spring of 1984 I decided that I would try to write a *New Yorker* profile of Bell whom I had known since he first went to CERN in 1960. We had a new editor at the *New Yorker*, Robert Gottlieb, who did not seem to have that much interest in science, but since I was going to CERN anyway on a leave there was not much to lose. Bell seemed agreeable and over some days I interviewed him on tape. Later I wrote my profile which was turned down. I published it in a 1991[17] collection *Quantum Profiles*. By the time the book came out John had died. He died on October 1 of 1990 of a cerebral aneurism. He had been nominated for a Nobel Prize which I think he would have won. He had also become something of a cult figure especially among New Age types who had no real understanding of what he had done. John seemed to accept all of this with a wry amusement. In 1979 he even attended a meeting organized by the Maharishi Mahesh Yogi, who had in fact been a physics major, which took place at the Maharishi university above Lake Lucerne. Bell told me that while he found the occasion rather absurd he liked the vegetarian meals. During my interviews I got the impression that none of the formulations of the quantum theory really satisfied him. I think the de Broglie-Bohm came closest although he was bothered by making it Lorentz invariant. He said that someday he might write a book about all of this. He never did.

---

[16] "Can Quantum Mechanical description of physical reality be considered complete?" A, Einstein, B.Podolsky and N. Rosen, Physical Review, **47**, 696 (1935.)

                Jeremy Bernstein

Appendix: Spinning[18]

      In the body of the text I mentioned some of the consequences of rotating the Stern-Gerlach magnets. In this appendix I want to fill in the details. We imagine first performing measurements of the spin along the z-axis when the particles are moving in the y direction. We then rotate the magnet through an angle $\theta$ in xz plane. The Pauli matrix which was

---

[17] Princeton University Press, Princeton
[18] I am very grateful to David Mermin for the critical remarks on an earlier draft that inspired me to write this appendix. I am also grateful to Elihu Abrahams for a critical reading of this draft,

$\begin{pmatrix} 1 & 0 \\ 0 & -1 \end{pmatrix}$ is in the new system $\begin{pmatrix} \cos(\theta) & \sin(\theta) \\ \sin(\theta) & -\cos(\theta) \end{pmatrix}$. This matrix has the eigen vectors $\begin{pmatrix} \cos(\theta/2) \\ \sin(\theta/2) \end{pmatrix}$ and $\begin{pmatrix} -\sin(\theta/2) \\ \cos(\theta/2) \end{pmatrix}$. We can expand the vector $\begin{pmatrix} 1 \\ 0 \end{pmatrix}$ is this basis and write

$\begin{pmatrix} 1 \\ 0 \end{pmatrix} = a_+ \begin{pmatrix} \cos(\theta/2) \\ \sin(\theta/2) \end{pmatrix} + a_- \begin{pmatrix} -\sin(\theta/2) \\ \cos(\theta/2) \end{pmatrix}$. Which implies that

$a_+$= cos(θ/2) and $a_-$=-sin(θ/2). This means that the probability of finding the spin up in rotated magnet is cos(θ/2)² while the probability of finding spin down is sin(θ/2)². Hence with the entangled singlet particles if I say measure spin down (or up) in one magnet then the probability of measuring the same result in the rotated magnet is sin(θ/2)² while the probability of measuring the opposite spin is cos(θ/2)². Thus the quantum mechanical correlation is given by

sin(θ/2)²- cos(θ/2)²=-cos(θ).

Can we program the robots to reproduce this? There is no problem programming a robot when it finds the rotated magnet to alter its trajectory so that the two spins are aligned

sin(θ/2)² fraction of the time agreeing with quantum mechanics. But if both magnets are rotated in opposite directions by the same angle then the robots will alter their trajectories so that agreement occurs 2 sin(θ/2)² of the time. But the quantum prediction is that agreement in this case occurs sin(θ)² percent of the time. In the range 0< θ<π/2

sin(θ)²>2 sin(θ/2)² as the figure below shows. This is Bell's inequality in this simple case.

The blue line is the plot for sin(θ)² and the red line is the plot for 2 sin(θ/2)².

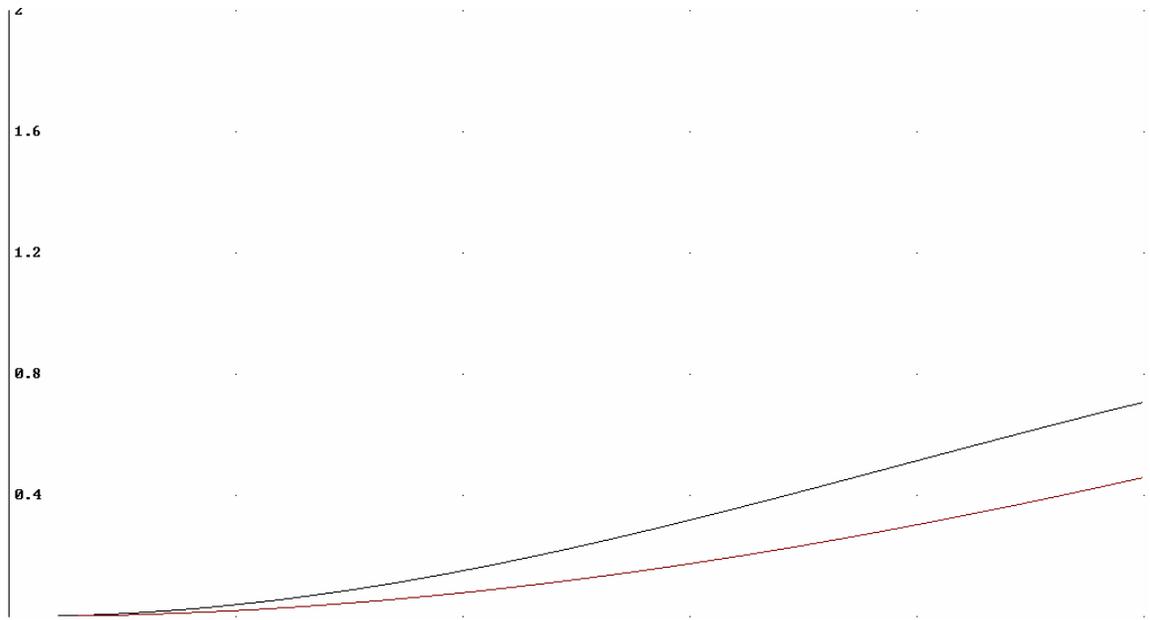